%% file: manuscript.tex
\documentclass[runningheads]{llncs}
\usepackage{graphicx}
\usepackage{todonotes}
\usepackage[T1]{fontenc}
\usepackage{import}
\usepackage{multirow}
\usepackage{url}
\usepackage{hyperref}

\begin{document}
\title{Preserving the Artifacts of the Early Digital Era: A Study of What, Why and How?}

\titlerunning{Preserving the Artifacts of the Early Digital Era}
%

\hypersetup{
    pdftitle={Preserving the Artifacts of the Early Digital Era: A Study of What, Why and How?},
    pdfsubject={Preserving the Artifacts of the Early Digital Era: A Study of What, Why and How?},
    pdfauthor={Maciej Grzeszczuk, Kinga Skorupska},
    pdfkeywords={computing history, retrocomputing, magnetic media,  early digital artifacts, demoscene},
    linkcolor={red!50!black},
    citecolor={blue!50!black},
    urlcolor={blue!80!black},
    hidelinks
}

\author{
Maciej~Grzeszczuk\inst{1, 2}\orcidID{0000-0002-9840-3398}\and
Kinga~Skorupska\inst{1}\orcidID{0000-0002-9005-0348} 
}
\authorrunning{M. Grzeszczuk and K. Skorupska}
%
\institute{Polish-Japanese Academy of Information Technology\\
\and
The Foundation for the History of Home Computers\\
\email{maciej.grzeszczuk@fhkd.pl}
}

\maketitle              
\vspace{-12pt}
\begin{abstract}

In this article, we report the pilot results of a survey study (N=1036) related to social attitudes towards the early digital heritage. On the basis of the answers, we consider what constitutes early digital artifacts (EDA) and outline how knowledge about them can be useful. We explore attitudes toward the historical and cultural importance of various EDAs and chart the surveyed requirements for their successful and sustainable preservation for current and future generations. 

\keywords{computing history \and retrocomputing \and magnetic media \and early digital artifacts \and demoscene}

\end{abstract}
\section{Introduction}
Visiting a museum, we may come across a receipt from a shoemaker from 100 years ago. At the time it was issued, it was a common thing of no special significance.
However, it escaped oblivion in the wars that swept through Europe and dodged the landfill. On its own, this just makes it a lucky receipt. But juxtaposed with other testimonies of that time, a photograph of the street where the workshop was located, a letter by a resident who used the services of a local craftsman, or even a distinctive shoe print in the hot asphalt of the street, it gives us enough to attempt to recreate the unique historical context. These testimonies can be put together because the media containing them have survived to be read and interpreted. Testimonies on paper, although seemingly fragile, can now be recovered even from charred 2000-year-old scrolls \cite{Parsons2023EduceLabScrollsVR}. Counterintuitively, the recovery and preservation of content from more recent media also constitutes a significant challenge.
The gradual popularization of computers in the second half of the 20th century created many new, previously unknown possibilities and applications, but it came with new threats to the continuity of historical records. Hidden in the shadow of this mainstream culture is a complex patchwork of everyday human activity, motivated by individual creativity, all in the context of the prevailing socioeconomic circumstances of times and places. These stories exist on unapproachable magnetic media, often unread for decades, home-built hardware, notebooks filled with mnemonics and in the minds of people who drove and witnessed this revolution, who are still alive today. This gives hope for the success of this last call for coordinated actions aimed at preserving the remaining data from that time. In this article, we report the pilot results of a survey study (N=1036) related to social attitudes toward the early digital heritage. We explore views on the historical and cultural importance of various early digital artifacts (EDA) and chart the surveyed requirements for their successful and sustainable preservation for current and future generations. 

\section{Related Work}

Historians work with objects to create a narrative that allows for their interpretation \cite{song2023,merriman1991beyond}. But since the 1970s these objects increasingly lost their physical form, moving into the digital sphere with the rise of early computers. Thus, digital artifacts also become a research subject, whether in the context of computers \cite{weber2013,sola1997essays}, software recovery \cite{vries2016,stachniak2019recovery} or early networking and human interactions \cite{stachniak2014eds,bates1993}. A great example of these complex computer-driven activities is the demoscene (or "the scene"), a phenomenon that evolved from the crack intros\footnote{Crack into was a short program, added to be started before the game which copy protection had been removed (cracked). It was a showcase of programming skills mixed with a social interaction tool - short text scrolls carried various messages.} and swap meetings \cite{pigulak2023} where people exchanged not only programs, but also knowledge \cite{roeder2022}, ideas, and artistic creations \cite{rubio2020}, connecting enthusiasts despite the barriers imposed by political systems or economic situation\footnote{Wasiak P., Szałankiewicz Ł., Lichnerowicz A.: "Polska demoscena jako wspólnota", ISBN: 978-83-969945-0-9, \url{https://kskpd.pl/album2023}} \cite{pigulak2023}. As time has passed, it naturally incorporated retrocomputing - a fascination with outdated computer hardware and software. While retrocomputing is also the subject of researchers' attention \cite{behrendtz2021}, it has sprouted communities that deal with the analysis, preservation, and popularization of remnants of digital culture independently of the scientific world. In 2004, Marcin "Stryker" Wojciechowski started an initiative to collect unofficial cassette releases of the software\footnote{Homepage of Cas Archive can be found at: \url{http://cas-archive.pigwa.net}}, Chilean community preserves remnants of the 8-bit Atari era\footnote{MUSEO DE CASETES ATARI: 
\url{https://retrogames.cl/museoataricas.html}
} and AtariOnline.pl\footnote{Link to file library: 
\url{https://www.atarionline.pl/v01/index.php?ct=katalog}} has slightly more than 2000 mixed origin cassette releases in their archives. Sebastian "Seban" Igielski analyzes Atari tape recorders for modifications of the transmission method\footnote{Such were widespread in countries where economic constraints did not allow having a disk drive. Popular examples like Turbo 2000 KSO, Turbo Blizzard, or Atari Super Turbo could effectively accelerate the tape recorder up to 8 times.} and publishes decoded files, photos, and schematics along with an analysis of their operation back to the community\footnote{The atari.area forum thread: \url{http://www.atari.org.pl/forum/viewtopic.php?id=12348}. Cassettes repository: \url{http://seban.pigwa.net/uicr0bee/}}. Such preservation efforts occur within the scene and among the retrocomputing "nerds"; however, with this research, we examine the perceived importance and purpose of early digital heritage preservation among the broader community and the public.

\begin{figure}
\centering
\includegraphics[width=\textwidth]{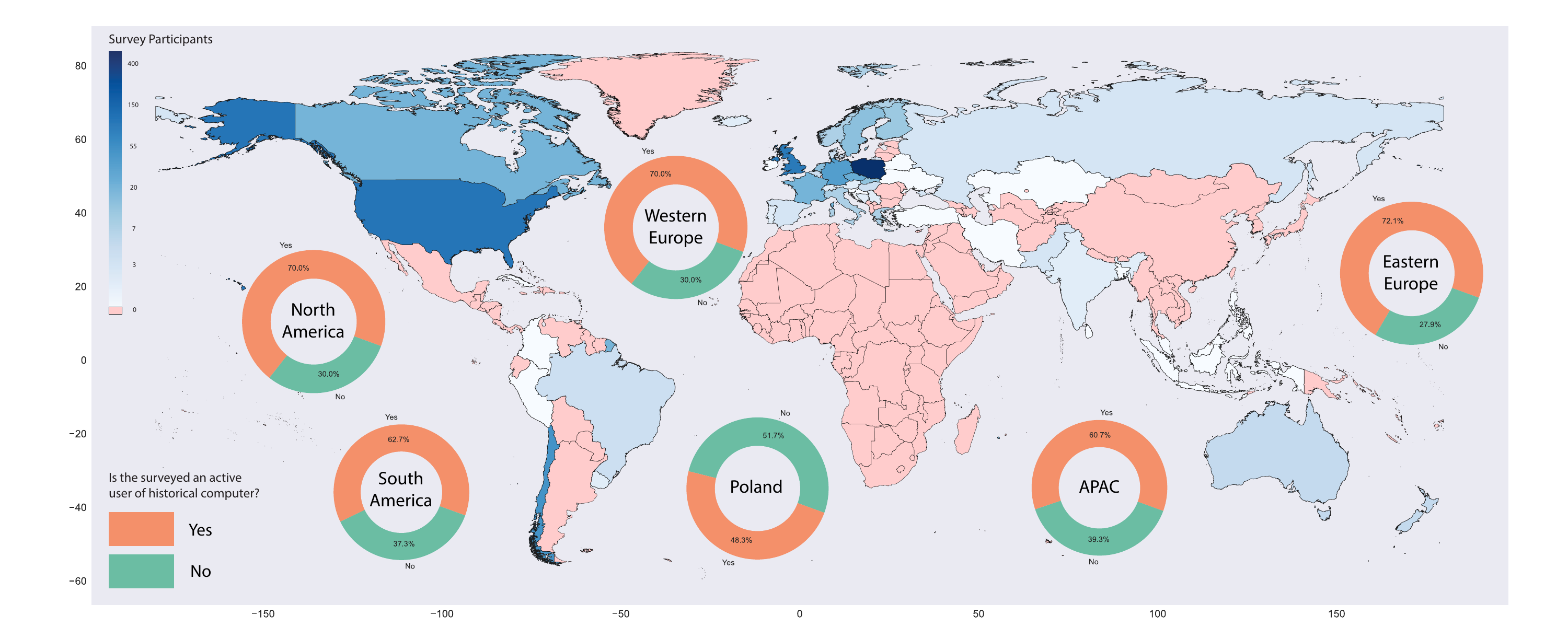}
\caption{A world map showing the territorial scope of our survey. Salmon color means no respondents from the country. The intensity of the blue indicates the number of participants on a logarithmic scale (legend in the upper left corner). The pie charts indicate share of active users of historical computers.}
\label{world}
\end{figure}

\section{Methods}

In March 2023, we conducted a survey on the importance of the remnants of the early home computing era. We wanted honest and open expression, including critical, drawing from the concepts of action research \cite{bradbury2003action}. The first batch of 36 was pen and paper. It included a briefing to verify whether the survey was clear. We then launched it online, gathering more than 500 responses in the first week, for a total of 1036 responses in 3 months it was active. We prepared it in English, Polish, and Spanish and were reaching out through direct messaging and social media, including diverse hobby/professional groups, retrocomputing discussion forums, Facebook, Reddit, and Discord groups. We encouraged to share it with family, spouses, and friends regardless of the age or experience in computing.

\section{Results and Discussion}

\subsection{Survey Demographics}
10.9\% of surveyed were female, 87.5\% male, 1\% did not disclose sex and 0.7\% indicated "other". 63\% of the respondents were between 36 and 49, aged between 50 and 59 constituted 24.8\%, in the age range of 27-35 there were 5.7\% of the participants, over 60 years of age were declared by 4.7\%, people between 18 and 26 years of age constituted 1. 6\% and we had 1 person under 18 years of age (with parental consent). We have reached 49 countries and five continents (see Figure \ref{world}). Of 1036 participants 595 were historical computer users.

\subsection{What is an Artifact of the Early Digital Age?}

\begin{figure}
\centering
\includegraphics[width=0.9\textwidth]{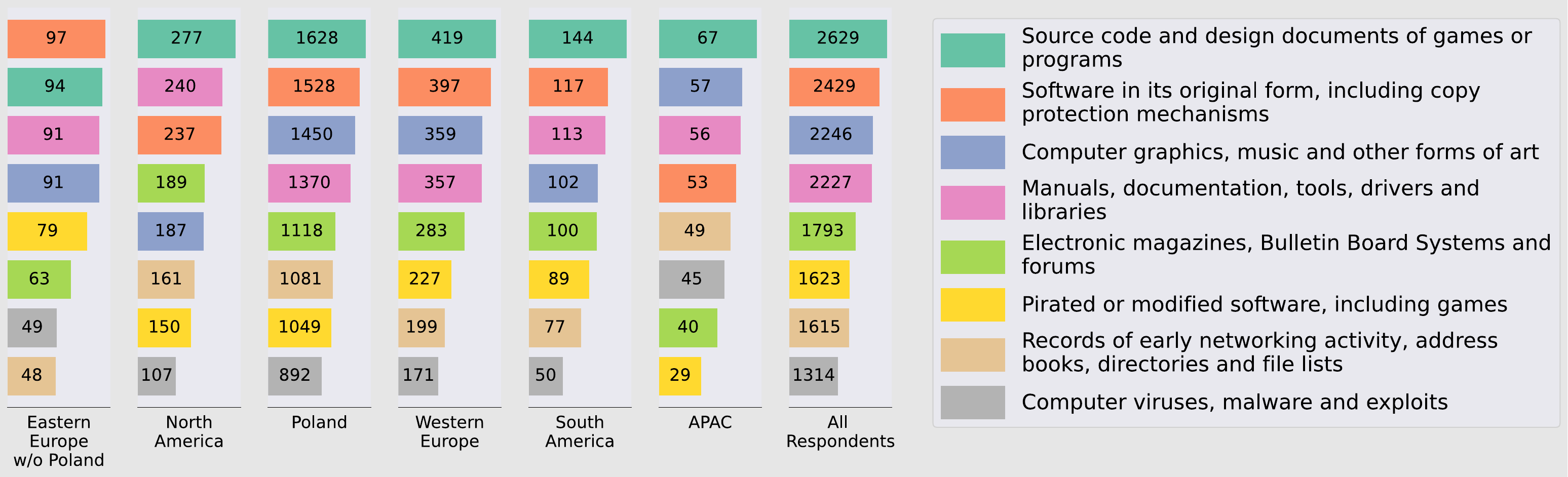}
\caption{Cumulative importance scores for a given EDA type. The one selected as the most important received a score of 8, the least important a score of 1.}
\label{edas}
\vspace{-0.6cm}
\end{figure}

We asked non-historical computer users to reorder a randomized list of remnants of the digital era based on their importance. Almost unanimously, they placed the source code and design documents at the top (see Figure \ref{edas}). With minor regional differences, the podium also included original software, as well as music, graphics, and other forms of artistic expression. Although we did not include equipment on the list for this question so as not to distort the results with its material value, the respondents drew attention to its importance in the comments.

\subsection{Why to Preserve Artifacts of the Early Digital Age?}

EDAs are believed to be records carrying a broader context than the history of the systems it is represented in (247 respondents; see Table \ref{tab:purpose} for this and the following annotations). They can help understand the evolution of technology and HCI interaction (204) and be used to learn computing (179), as "historic systems and software are hugely more illustrative than modern comparable systems, being restricted to a scope that is a lot easier to understand". Algorithms and interaction paradigms popular in the past, but since lost to history, may have their renaissance today (90). In the same vein, some problems and errors encountered in the past may serve as lessons, so they are not repeated today (47). The benefits are not limited to early computing education, as experience in finding creative solutions in resource-constrained environments can help create a better code on modern platforms (130), as one "thinks about the impact of code quality on infrastructure". One respondent eloquently puts it: "Understanding physical computers, physical networks will help understand our virtual world. Virtual worlds are beautiful, but they are still zeroes and ones, an assembly language, cpu language, machine language, electricity, logic gates, transistors. Your beautiful worlds you build in minecraft are still simply magnetic data somewhere on a harddisk in a faraway data centre.". Another reason given is that early computing allows for a more DIY element, as the tools are simpler and, thus, are easier to modify -- allowing enthusiasts to take full ownership and control of the devices and software -- unlike their modern, closed, equivalents.

\begin{table}
    \centering
    \tiny
    \caption{Responses for question "How can knowledge about historical computers and software be useful to us today or in the future?", presented by region, for for users and non-users of historical computers regardless of region, and also as a percentage of all responses.}
    \import{./tables/}{purpose.tex}
    \label{tab:purpose}
\end{table}

Although there were skeptical voices about the sense of preserving EDAs (12 respondents) and the belief that they had no practical application (49), an even larger group of 64 was convinced that artifacts can be used to inspire and educate next generations. One of the youth is enthusiastically recalling an event during which the atmosphere of a computer fair from the 1980s was recreated: "I like old stories related to historical computers. I wish there [still] was a hacker with a cassette [stand]. He had a very funny book. "Space", "Nice Maze", "Collecting eggs and escaping from chickens"\footnote{In the times of stalls selling pirated software in Eastern Europe, the purchasing decision was made as a result of browsing through primitive, photocopied lists of programs, where each title was described with a single sentence, such as: "You are in a pyramid, collecting diamonds".}. Therefore, it seems important to choose the right presentation formula - the opportunity to see historical technology in action was important to 78 respondents and is also supported in the literature\cite{weber2013}. 

\subsection{How to Preserve Artifacts of the Early Digital Age?}

\begin{figure}
\centering
\includegraphics[width=\textwidth]{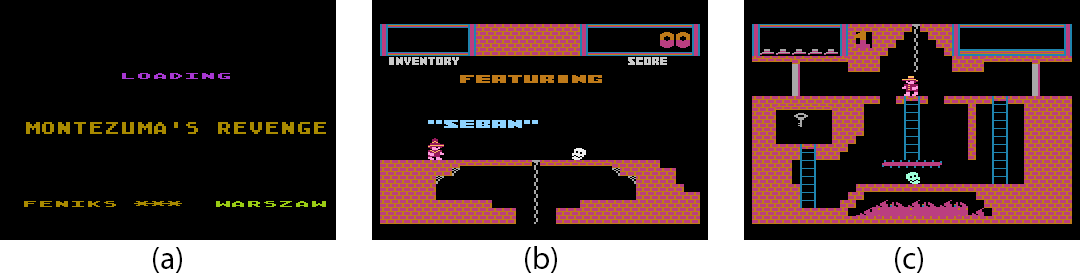}
\caption{Screenshots of multiple versions of the popular Atari XL/XE game Montezuma's Revenge: (a) Loader with an ad of the pirate stand operating in DH Feniks, Warsaw, Poland in late 1980s; (b) Modified version of the leaked demo game with signature of the cracker: "Seban" instead of "Pedro"; (c) The official release of the game.}
\label{montezumy}
\end{figure}

Our respondents see that a "community" (57 responses, see Table \ref{tab:methods} for this and the following annotations) and "organized effort" (133) are required to preserve EDA. Apart from that they mention educational actions (78) and campaigns (66) to make people "understand that their artifacts could be significant now or beyond their lifetime, and to ensure they're passed to relatives or people/bodies (museums etc), who care about their future.". This is valuable, but may be too late, as "oral histories", "memories", and experienced "emotions" (13) may be lost at that point\footnote{A perfect example is an interview with Robert Jaeger, author of Montezuma's Revenge (see Figure \ref{montezumy}). He mentions that the game that circulated in a seemingly more complete form, with an animated intro and music, was the result of the leak. The release version had been truncated for commercial reasons and looked like preliminary in comparison (hence it was commonly known as Preliminary Monty). The link to the interview: \url{https://www.youtube.com/watch?v=KgerqFv6tNo}.}, just as documentation (58) about the artifacts kept. As one respondent insightfully states: "The issue with software and old computers is that merely looking at an inert device is only experiencing a tenth of the significance or context of this device as a whole. [...] As old operators die, new ones will have to replace them"

Respondents think that we need to establish "libraries", where they can be "preserved for eternity", but they note that "legislation is not up to date with this" (24). One suggests that "software should become open-source the moment it becomes abandonware. (...) A community will have much more capability to preserve it rather than its creator who's no longer interested in it.".

Apart from moving them from "rotting media" (252), working hardware and emulators are also necessary (106). Some respondents mention the challenges of having enough resources (64) and time (27) for the job, and also to check them periodically (161). Still, preservation limited to physical spaces is risky, as the "destruction of a computer museum at the beginning of the war in Ukraine" shows. Online archives (106), "such as www.pouet.net, www.archive.org" are promising,  especially if they would facilitate a "coordinated effort to catalog and digitally store software"\footnote{The lack of a unified system used in practice is a problem for existing initiatives. Most different versions of the same program in the atarionline.pl library are distinguished only by a consecutive number, without any indication of their differences, not to mention the potential stories associated with them.} with "appropriate metadata, similar to FAIR principles" (https://www.go-fair.org/fair-principles/). Paper-based material also needs to be digitized to be stored in the cloud (58). People should be allowed to upload their materials to a public database with clear tags and descriptions -- that would allow them to also add and describe their own prototypes and creative works, previously unavailable online. As one respondent notes, currently such artifacts exist and are dispersed in diverse collections, but they are not cataloged, sometimes no longer accessible. Finally, "the more people have access to these artifacts, the more likely they’ll be preserved because there will be people saving them to their devices.".

\begin{table}
    \centering
    \tiny
    \caption{Responses for question "What would it take to successfully preserve these digital artifacts?", presented by region, for for users and non-users of historical computers regardless of region, and also as a percentage of all responses.}
    \import{./tables/}{methods.tex}
    \label{tab:methods}
\end{table}

\section{Conclusion and Future Work}

Overall, a great majority of our respondents saw a reason to preserve early digital artifacts, or EDAs (Table \ref{tab:purpose}), mostly as historical and educational sources. Apart from digital artifacts they considered important, such as source code, design docs, original software and art forms (Figure \ref{edas}) they mentioned the need to preserve hardware. As for EDAs, the efforts ought to focus on digitizing, either for online archives or cultural institutions (Table \ref{tab:methods}). The existence of huge online archival initiatives may discourage people from backing up individual collections, as they may feel those are already digitized. While this may be true for original editions of mass-released titles, it is not so in the case of bootleg recordings or works of art, potentially still in the hands of the community, waiting to be preserved. It is the existence of the unregulated scene which is an important testament to the ingenuity of people faced with economic inequality and scarcity. These artifacts not only tell individual stories of creativity, but also help recreate the rich socioeconomic context of the early computing era in different regions. This context is ever incomplete, and it becomes harder to recreate as more artifacts degrade and memories fade \cite{behrendtz2021,vries2016}. The preservation task is even more difficult as it is often performed as uncoordinated, individual actions, without systemic and legislative support -- and the goal of our future work is to address these challenges. We need artifacts standing testimony to the everyday reality of the past. The ones that gave rise to the early HCI community; the people who produced code written within the brutal constraints of early hardware we can use to educate future generations \cite{sola1997essays}. Without this context, we cannot fully appreciate how far we have come, what it took to arrive here, and what roads were not taken.

\newpage

\section*{Acknowledgments}

The authors express their gratitude to all survey respondents for sharing their personal views and experiences in this study. We also want to recognize the massive effort of the community of enthusiasts who tackle collection, maintenance, and preservation tasks, often without much systemic or institutional support.

%
%
%
\bibliographystyle{splncs04}
\bibliography{bibliography}

\end{document}

%% file: tables/purpose.tex
\begin{tabular}{l|cccccc|ccc}
                                           & \multicolumn{6}{c|}{Responses by Region}                                                                                                                                                                                                                              & \multicolumn{3}{c}{All Regions}     \\
                                           & \rotatebox{90}{\begin{tabular}[c]{@{}c@{}}Eastern Europe \\ w/o Poland\end{tabular}} & \rotatebox{90}{\begin{tabular}[c]{@{}c@{}}North America\end{tabular}} & \rotatebox{90}{Poland} & \rotatebox{90}{\begin{tabular}[c]{@{}c@{}}Western Europe\end{tabular}} & \rotatebox{90}{\begin{tabular}[c]{@{}c@{}}South America\end{tabular}} & \rotatebox{90}{APAC} & \rotatebox{90}{Users}        & \rotatebox{90}{Non-Users}  &
                                           \rotatebox{90}{All Responses} \\ \hline
As historical records in broader context   & 14                                                                  & 23                                                       & 151    & 37                                                        & 9                                                        & 13   & 137   & 110       & 27.1\%          \\
To explore history of technology           & 12                                                                  & 35                                                       & 87     & 49                                                        & 14                                                       & 7    & 119   & 85        & 22.4\%          \\
To learn computing on simple models        & 14                                                                  & 22                                                       & 77     & 56                                                        & 10                                                       & 0    & 125   & 54        & 19.6\%          \\
To create better code, efficient solutions & 13                                                                  & 20                                                       & 56     & 30                                                        & 10                                                       & 1    & 88    & 42        & 14.3\%          \\
To study old design, to remake old ideas   & 6                                                                   & 18                                                       & 43     & 13                                                        & 10                                                       & 0    & 56    & 34        & 9.9\%           \\
To develop and inspire next generations    & 3                                                                   & 6                                                        & 38     & 9                                                         & 4                                                        & 4    & 42    & 22        & 7.0\%           \\
To appreciate where we are now             & 4                                                                   & 11                                                       & 18     & 20                                                        & 5                                                        & 1    & 35    & 24        & 6.5\%           \\
Nostalgia                                  & 3                                                                   & 4                                                        & 41     & 9                                                         & 0                                                        & 0    & 24    & 33        & 6.2\%           \\
Fun, Hobby, Tinkering, Art                 & 3                                                                   & 8                                                        & 38     & 6                                                         & 1                                                        & 0    & 39    & 17        & 6.1\%           \\
To foresee future trends                   & 3                                                                   & 11                                                       & 27     & 9                                                         & 4                                                        & 2    & 28    & 28        & 6.1\%           \\
It has no use                              & 2                                                                   & 4                                                        & 31     & 7                                                         & 1                                                        & 4    & 30    & 19        & 5.4\%           \\
It is just a curiosity                     & 1                                                                   & 2                                                        & 38     & 6                                                         & 1                                                        & 0    & 18    & 30        & 5.3\%           \\
To learn from past mistakes                & 3                                                                   & 11                                                       & 20     & 11                                                        & 2                                                        & 0    & 21    & 26        & 5.2\%           \\
To be able to recover civilisation         & 0                                                                   & 0                                                        & 14     & 2                                                         & 1                                                        & 0    & 7     & 10        & 1.9\%           \\
Legacy system continuity                   & 0                                                                   & 1                                                        & 7      & 5                                                         & 1                                                        & 1    & 12    & 3         & 1.6\%           \\ \hline
All Responses                              & 56                                                                  & 114                                                      & 474    & 190                                                       & 53                                                       & 25   & 537   & 375       & 100\%          
\end{tabular}

%% file: tables/methods.tex
\begin{tabular}{l|cccccc|ccc}
                                           & \multicolumn{6}{c|}{Responses by Region}                                                                                                                                                                                                                              & \multicolumn{3}{c}{All Regions}     \\
                                           & \rotatebox{90}{\begin{tabular}[c]{@{}c@{}}Eastern Europe \\ w/o Poland\end{tabular}} & \rotatebox{90}{\begin{tabular}[c]{@{}c@{}}North America\end{tabular}} & \rotatebox{90}{Poland} & \rotatebox{90}{\begin{tabular}[c]{@{}c@{}}Western Europe\end{tabular}} & \rotatebox{90}{\begin{tabular}[c]{@{}c@{}}South America\end{tabular}} & \rotatebox{90}{APAC} & \rotatebox{90}{Users}        & \rotatebox{90}{Non-Users}  &
                                           \rotatebox{90}{All Responses} \\ \hline
Digitizing and backing up physical media & 8                                                                   & 33                                                       & 153    & 50                                                        & 5                                                        & 3    & 161   & 91        & 29.7\%          \\
Regular use and maintenance              & 6                                                                   & 14                                                       & 96     & 25                                                        & 14                                                       & 6    & 106   & 55        & 19.0\%          \\
Museums and institutions                 & 4                                                                   & 7                                                        & 85     & 28                                                        & 5                                                        & 4    & 59    & 74        & 15.7\%          \\
Controlled storage conditions            & 3                                                                   & 17                                                       & 77     & 21                                                        & 10                                                       & 3    & 76    & 55        & 15.4\%          \\
Hardware, spare parts or emulators       & 7                                                                   & 16                                                       & 36     & 42                                                        & 5                                                        & 0    & 72    & 34        & 12.5\%          \\
On-line archives and databases           & 6                                                                   & 17                                                       & 52     & 26                                                        & 2                                                        & 3    & 65    & 41        & 12.5\%          \\
Demonstrations and engagement            & 3                                                                   & 4                                                        & 53     & 9                                                         & 5                                                        & 4    & 55    & 23        & 9.2\%           \\
Raising public awareness of the topic    & 6                                                                   & 3                                                        & 42     & 8                                                         & 4                                                        & 3    & 43    & 23        & 7.8\%           \\
Knowledge, skills and research           & 6                                                                   & 5                                                        & 24     & 25                                                        & 1                                                        & 3    & 45    & 19        & 7.5\%           \\
Money and resources                      & 7                                                                   & 14                                                       & 14     & 19                                                        & 7                                                        & 3    & 46    & 18        & 7.5\%           \\
Digitizing or recreating documentation   & 1                                                                   & 12                                                       & 29     & 13                                                        & 1                                                        & 2    & 44    & 14        & 6.8\%           \\
Building and engaging community          & 2                                                                   & 5                                                        & 27     & 15                                                        & 7                                                        & 1    & 33    & 24        & 6.7\%           \\
Passion, drive and motivation            & 2                                                                   & 8                                                        & 9      & 19                                                        & 12                                                       & 3    & 33    & 20        & 6.2\%           \\
Descriptions, metadata and cataloging    & 1                                                                   & 1                                                        & 22     & 7                                                         & 0                                                        & 1    & 21    & 11        & 3.8\%           \\
Building collections                     & 1                                                                   & 2                                                        & 23     & 5                                                         & 0                                                        & 0    & 18    & 13        & 3.7\%           \\
Time                                     & 2                                                                   & 7                                                        & 2      & 12                                                        & 4                                                        & 0    & 18    & 9         & 3.2\%           \\
Legislative changes                      & 1                                                                   & 4                                                        & 8      & 6                                                         & 2                                                        & 3    & 13    & 11        & 2.8\%           \\
It is already preserved                  & 2                                                                   & 6                                                        & 6      & 8                                                         & 1                                                        & 0    & 19    & 4         & 2.7\%           \\
Tools                                    & 1                                                                   & 8                                                        & 2      & 9                                                         & 0                                                        & 2    & 19    & 3         & 2.6\%           \\
Gathering personal histories             & 0                                                                   & 3                                                        & 8      & 1                                                         & 0                                                        & 1    & 7     & 6         & 1.5\%           \\
There is no need to do it                & 0                                                                   & 0                                                        & 9      & 3                                                         & 0                                                        & 0    & 7     & 5         & 1.4\%           \\ \hline
All Respondents                          & 40                                                                  & 100                                                      & 446    & 185                                                       & 55                                                       & 23   & 510   & 339       & 100\%          
\end{tabular}